\documentclass[a4paper,prl,twocolumn]{revtex4}
\usepackage{amsmath}
\begin{document}
\title{Strong correspondence principle for joint measurement of conjugate observables} 
\author{A. Di Lorenzo}
\affiliation{Instituto de F\'{\i}sica, Universidade Federal de Uberl\^{a}ndia, 38400-902 Uberl\^{a}ndia MG, Brasil}
\begin{abstract}
It is demonstrated that the the statistics for a joint measurement of two conjugate variables in Quantum Mechanics 
are expressed through an equation identical to the classical one, provided that joint classical probabilities are substituted 
by Wigner functions and that the interaction between system and detectors is accounted for. 
This constitutes an extension of Ehrenfest correspondence principle, and it is thereby dubbed strong correspondence 
principle. 
Furthermore, it is proved that 
the detectors provide an additive term to all the cumulants, and that if they are prepared in a Gaussian state they 
only contribute to the first and second cumulant. 
\end{abstract}

\maketitle

Simultaneous knowledge of two conjugate observables is a discriminating feature 
between Classical and Quantum Mechanics: it is possible, at least in principle, in the former, and impossible in the latter. 
For Quantum Mechanics this should be interpreted as the non-feasibility of knowing with arbitrary precision both 
observables, not as the impossibility of realizing joint measurements. 
Indeed, in a pioneering work, Arthurs and Kelly \cite{Arthurs65} 
demonstrated that a scheme exists for joint measurement of position and momentum of a particle, albeit at the cost 
of sacrificing the precision of both measurements. 
In \cite{Arthurs65} very specific 
hypotheses were made about the initial preparation of the detectors. 
Later on, these assumptions were relaxed \cite{Braunstein91,Stenholm92}, and the connection of the joint probability to the Husimi $Q$ function \cite{Husimi} was found. 
Remarkably, joint measurements of conjugate observables have been realized in optics, either through 
multiport techniques \cite{Walker} or homodyne detection \cite{Noh}. 
Furthermore, it was demonstrated\cite{Arthurs88} that in the joint measurement the product of the uncertainty 
of the conjugate variables is not less than $\hbar$, twice as much as the limit of the uncertainty principle, which 
applies instead to separate measurements of position and momentum. 
This result was followed by several papers discussing 
the interpretation of the uncertainty relations starting from the 90s \cite{uncertainty90s} up to recent years \cite{uncertainty00s}. 
On another front, the conditional state of the system after the measurement was studied \cite{Arthurs65,She66}. 
Ref. \cite{She66}, however, assumed that the statistics is fully determined by the 
observed values and the associated spreads, from which a gaussian function was 
constructed. This in general is not correct, since the statistics of the outcomes is not 
always gaussian, and higher order cumulants are needed to characterize it. 

In the present work I shall prove that the formula giving the probability density for the output of the detectors is 
identical to the one which is obtained in the classical case, 
provided that joint probabilities for the system observables and for the detectors observables are 
substituted by the Wigner quasi-probabilities functions. 
This result goes beyond Ehrenfest theorem, which applies to average values only. 
Analogously, I demonstrate that, considering the conditional state of the system given 
the outcomes, the formula which expresses such a state in terms of Wigner function is identical 
to the classical equation. Also, the more general expression for the conditional state of the system is derived, and it differs from the one surmised in \cite{She66}.

I consider detection of two conjugate variables, denoted $\Hat{Q},\Hat{K}$, with units such that $[\Hat{Q},\Hat{K}]=i$. 
The interaction between detectors and system is written 
\begin{equation}\label{eq:hint}
H_{int}=-\delta(t-t_0) \sum_{A=Q,K} \lambda_A \Hat{\phi}_A \Hat{A}, 
\end{equation}
with $\Hat{\phi}_A$ an operator on the $A$-th detector Hilbert space, and $\Hat{A}$ the observables of the 
system, $\lambda_A$ coupling constants and $t_0$ time of the measurement. 
This coupling corresponds to the standard von Neumann detection scheme \cite{vonneumann}, 
duplicated to allow joint measurement 
of non-commuting observables. As such it is known as the Arthurs-Kelly scheme, after the authors who 
introduced it for the first time \cite{Arthurs65}.
For a single measurement (e.g. $\lambda_K=0$), in order to have an ideal measurement, 
(i) initially, the density matrix of detector and system is $\rho=\rho_S \otimes \rho_A$;
(ii) the detector must be prepared in a sharp state $\rho_{det}(I_A,I_A')=|I_A=0\rangle\langle I_A=0|$, peaked around 
$I_A=0$. 

Indeed, under assumptions (i) and (ii), after the interaction with the system 
the probability distribution for the variable $A$ is 
\begin{equation}\label{eq:onemeasprob}
\Pi(I_A|\Hat{A})=\int d\mu(a) \delta(I_A-\lambda_A a) \langle a | \Hat{\rho}_s(t_0^-)| a \rangle, 
\end{equation}
$a, |a\rangle$ being eigenvalues and eigenstates of $\Hat{A}$, and $\mu(a)$ a measure.  
Eq.~\eqref{eq:onemeasprob} 
implies that in any single measurement $I_A$ will be found to have one of the values $\lambda_A a$. 
We can relax hypothesis (ii) by requiring, more realistically, 
that (ii$'$) initially the detector is prepared in a state 
$\rho_{A}(I_A,I_A')$, with $\Pi(I_A):=\rho_{A}(I_A,I_A)$ peaked around $I_A=0$.  
Then Eq.~\eqref{eq:onemeasprob} becomes the convolution
\begin{equation}\label{eq:onemeasprob2}
\Pi(I_A|\Hat{A})=\int d\mu(a)  
\Pi_{A}(I_A-\lambda_A a) \langle a | \Hat{\rho}_S(t_0^-)| a \rangle .
\end{equation}
In terms of characteristic functions
\begin{equation}\label{eq:onemeascharfunc}
Z_{A}(\chi_A;t_0^+)= Z_S(\chi_A;t_0^-) Z_{A}(\chi_A;t_0^-).
\end{equation}
Actually, one can consider a finite-time measurement, i.e. substitute the delta-function with a regular function $g(t)$, 
vanishing outside a finite time window, under the additional assumptions: 
(iii) the Hamiltonian of the detector depends only on $\Hat{I}_A$, 
the variables conjugated to $\Hat{\phi}_A$, so that 
$\left[\Hat{\phi}_A,\Hat{I}_B\right]=\delta_{AB} i \hbar$; 
and (iv) the observable $\Hat{A}$ is conserved 
during the free evolution of the system (at least approximately). 
The measurement is then a non-demolition one \cite{Braginsky}. 
Since I am interested in measuring non-commuting observables, the only Hamiltonian conserving simultaneously 
$\Hat{Q}$ and $\Hat{K}$ would be the trivial zero Hamiltonian, thus I keep the instantaneous interaction, and make 
use of assumptions (i) and (ii$'$) throughout the rest of this paper. 
(In the case of joint measurement of  spin components, any spin-independent Hamiltonian conserves simultaneously all 
spin components, and then one could consider a finite duration measurement.)

Now, let us consider a simultaneous interaction with both detectors, i.e. $\lambda_A\neq 0,A=Q,K$. 
In the following I shall rescale the operators to eliminate the coupling constants and Planck's constant: 
$\lambda_A \Hat{\phi}_A/\hbar \to \Hat{\phi}_A$, $\Hat{I}_A/ \lambda_A \to\Hat{I}_A$. 
This way $I_Q,\Phi_K$ have the same dimensions as $Q$, while $I_K,\Phi_Q$ have the same dimensions as $K$. 
The formalism of quantum mechanics allows to ask what will be the probability density for $I_Q,I_K$. 
Indeed, applying Born's rule to the time evolution of the total density matrix, we have 
\begin{equation}
\Pi(\{I_A\}
)={\rm Tr}_S \left\{   \langle I_A| 
e^{i\sum_A \Hat{\phi}_A \Hat{A} }\rho^-  e^{-i\sum_A \Hat{\phi}_A \Hat{A} }
| I_A\rangle
\right\},
\end{equation}
with $\rho^-:=\rho_S\otimes \rho_{det}$, the density matrix of the system and detectors immediately before $t_0$.
After introducing twice the identity over the detectors Hilbert space in terms of eigenstates of $\Hat{\phi}_A$, 
we have the probability density
\begin{align}
\Pi(\{I_A\}
)=
&\int
\left(\prod_A \frac{d\chi_ A}{2\pi}\right)
\label{eq:multimeasprob}
 e^{-i\sum_B \chi_B I_B}
Z_{K,Q}(\{\chi_A\}),
\end{align}
where the generating function is
\begin{align}\nonumber
Z_{K,Q}(\{\chi_A\})=&
\int \left(\prod_A \frac{d\Phi_A}{2\pi}\right) K(\{\Phi_A\},\{\chi_A\})
\\
\label{eq:multimeaskern}
&\quad \times \rho_{det}(\{\Phi_A\!+\!\frac{\chi_A}{2},\Phi_A\!-\!\frac{\chi_A}{2}\})
 ,
\end{align}
with 
the kernel
\begin{align}\label{eq:multimeaskern1}
K(\{\Phi_A\},\{\chi_A\})=
{\rm Tr}_S \left\{
\Hat{V}_+ \rho_S\  V_-^\dagger
\right\},
\end{align}
and $\Hat{V}_\pm:=\exp{\left\{i\sum_A (\Phi_A\pm\chi_A/2) \Hat{A}\right\}}$.

Since the operators $\Hat{Q},\Hat{K}$ do not commute with each other, 
the kernel in Eq.~\eqref{eq:multimeaskern} will not be in general a function of $\chi_A$ only. 
Eq.~\eqref{eq:multimeaskern1} can be rewritten, 
after applying Baker-Campbell-Hausdorff formula, 
\begin{align}
&K
=
e^{i(\Phi_K \chi_Q+\chi_K \Phi_Q)/2}\
\label{eq:case1multimeaskern}
{\rm Tr}_S\! \left\{\Hat{V}_{Q+}
\Hat{V}_{K+}\rho_S
\Hat{V}_{K-}^\dagger \Hat{V}_{Q-}^\dagger 
\right\}, 
\end{align}
where we dropped the functional dependence for brevity and 
defined $\Hat{V}_{A\pm}:= \exp{\left\{i(\Phi_A\pm\chi_A/2) \Hat{A}\right\}}$

We can use the position eigenstates to express the trace, giving 
\begin{align}
K(\{\Phi_A\},\{\chi_A\})=&
\label{eq:case1multimeaskern2}
e^{i(\chi_K \Phi_Q-\Phi_K \chi_Q)/2}\ Z^W_S(\chi_K,\chi_Q)
,
\end{align}
with 
\begin{equation}
Z^W_S(q,k):=\widetilde{\Pi}^W_S(q,-k)=
 \int dQ
 e^{i k Q} \rho_S(Q+q/2,Q-q/2)
\end{equation}
the Fourier transform of Wigner quasi-probability $\Pi^W_S(K,Q)$. 
I changed the sign of the second variable so that $Z^W_S(0,\chi)$ is the generating function for the 
probability $\Pi_S(K)=\langle Q|\rho_S|Q\rangle$, 
and $Z^W_S(\chi,0)$ is the generating 
function for the probability $\check{\Pi}_S(K)=\langle K|\rho_S|K\rangle$. 
If the Wigner quasi-probability were positive defined, 
$Z^W_{S}(\chi_K,\chi_Q)$ would be the corresponding characteristic function.  
Since this is not in general the case, I shall call 
$Z^W_{S}$ a quasi-characteristic function. 

It follows readily from Eqs. \eqref{eq:multimeaskern} and \eqref{eq:case1multimeaskern2} 
that the characteristic function is 
\begin{align}
&\!\!Z(\chi_K,\chi_Q)\!=\!
\label{eq:charfuncqk}
Z^W_{S}\!(\chi_K,\chi_Q) 
Z^W_{det}(\chi_Q,\frac{\chi_K}{2};\chi_K,-\frac{\chi_Q}{2}) 
,
\end{align}
where 
\begin{align}
\nonumber
&Z^W_{det}(\chi_Q,i_Q;\chi_K,i_K):=\widetilde{\Pi}^W_{det}(\chi_Q,-i_Q;\chi_K,-i_K) \\ 
&=\int\! d\Phi_Q d\Phi_K e^{i\sum_A \Phi_A i_A} 
\langle \{\Phi_A+\chi_A/2\}|\rho_{det}|\{\Phi_A-\chi_A/2\}\rangle
\end{align}
is the quasi-characteristic function of the detectors.

Eq.~\eqref{eq:charfuncqk} 
should be contrasted with the case when first a measurement of $K$ is made, 
and shortly thereafter $Q$ is observed: 
\begin{align}
\label{eq:charfunckthenq}
Z_{<}(\chi_K,\chi_Q)=&
Z^W_{S}(\chi_K,\chi_Q) 
Z^W_{det}(\chi_Q,0;\chi_K,-\chi_Q) 
,
\end{align}
or viceversa 
\begin{align}
\label{eq:charfuncqthenk}
Z_{>	}(\chi_K,\chi_Q)
=& 
Z^W_{S}(\chi_K,\chi_Q) 
Z^W_{det}(\chi_Q,\chi_K;\chi_K,0) 
.
\end{align}

An important conclusion that can be drawn from Eq.~\eqref{eq:charfuncqk} is that 
the contribution of the detectors to all the cumulants (defined as logarithmic derivatives of $Z$ 
calculated at $\chi_Q=0,\chi_K=0$) is simply an additive term. In particular, if the detectors 
are prepared in a gaussian state, so is $Z_{det}$, and thus their contribution to the cumulants 
higher than the second one vanish.

The probability density is obtained by Fourier transforming Eq.~\eqref{eq:charfuncqk}, 
and it consists in a convolution of the Wigner quasi-probability densities: 
\begin{align}
\nonumber
\Pi(I_Q,I_K) =& \int dI_Q' dI_K' d\Phi_Q'd\Phi_K' \Pi^W_{det}(I_Q',\Phi_Q';I_K',\Phi_K') 
\\
&\times
\label{eq:probiqik}
\Pi^W_S(I_K\!-\!I_K'\!-\!\frac{\Phi_Q'}{2},I_Q\!-\!I_Q'\!+\!\frac{\Phi_K'}{2}) 
. 
\end{align}
A priori, it is not obvious that the convolution of two arbitrary quasi-probability functions presented in Eq.~\eqref{eq:probiqik} is positive-defined. However, $\Pi(I_Q,I_K)$ is positive by construction, thus an 
interesting mathematical corollary follows from the derivation presented above: 
given any two quasi-probability functions, their convolution as defined in Eq.~\eqref{eq:probiqik} gives a 
proper probability distribution. 
In particular, one can consider the detectors to be initially independent of one another, so that 
\begin{align}
\nonumber
\Pi(I_Q,I_K) =& \int dI_Q' dI_K' d\Phi_Q'd\Phi_K' 
\Pi^W_Q(I_Q',\Phi_Q') \Pi^W_K(I_K',\Phi_K') 
\\
&\times
\label{eq:probiqik2}
\Pi^W_S(I_K\!-\!I_K'\!-\!\frac{\Phi_Q'}{2},I_Q\!-\!I_Q'\!+\!\frac{\Phi_K'}{2}) 
. 
\end{align}
is positive definite for any three Wigner functions.

Furthermore, Eq.~\eqref{eq:probiqik} would have a simple interpretation if the 
Wigner quasi-probabilities were positive-defined: before the interaction, the observables of 
the detector $A$ possessed the values 
$I'_A$ and $\Phi'_A$  with probability $\Pi^W_{det}(I_Q',\Phi_Q';I_K',\Phi_K')$, 
and the $Q,K$ variables of the system 
had the values $Q',K'$, with probability $\Pi^W_S(Q',K')$. 
After the interaction, the value of $I_Q$ is shifted deterministically by $Q'-\Phi_K'/2$, and that of $I_K$ by $K'+\Phi_Q'/2$. 
 It is interesting to note that this is indeed the result one would obtain
in the classical case, if the interaction term is given by Eq.~\eqref{eq:hint}. 
After solving the classical Hamiltonian equations, the value of $I_Q,I_K$ immediately after the interaction are 
(primed quantities are calculated at $t_0^-=t_0-\varepsilon$, and unprimed ones at $t_0^+=t_0+\varepsilon$)
\begin{subequations}
\label{eq:classqk}
\begin{align}
\label{eq:classqka}
I_Q =& Q'+  I_Q'- \Phi_K'/2, \quad Q = Q' - \Phi_K',\\
\label{eq:classqkb}
I_K =& K' + I_K'+ \Phi_Q'/2, \quad K = K' + \Phi_Q',
\end{align}
\end{subequations}
from which Eq.~\eqref{eq:probiqik} readily follows.

It follows from Eq.~\eqref{eq:charfuncqk} that
\begin{subequations}\label{eq:averqk}
\begin{align}
\label{eq:averqka}
\langle I_Q \rangle =& \langle Q \rangle_S + \langle I_Q \rangle_Q - \langle \Phi_K \rangle_K/2,\\
\label{eq:averqkb}
\langle I_K \rangle =& \langle K \rangle_S + \langle I_K \rangle_K + \langle \Phi_Q \rangle_Q/2,
\end{align}
\end{subequations}
and 
\begin{subequations}\label{eq:spreadqk}
\begin{align}
\label{eq:spreadqka}
\langle \Delta I_Q^2 \rangle =& \langle \Delta \Hat{Q}^2 \rangle_S 
+ \langle \Delta \Hat{I}_Q^2 \rangle_Q 
+ \frac{1}{4}\langle \Delta \Hat{\Phi}_K^2 \rangle_K
+\langle \Delta \Hat{I}_Q \Delta \Hat{\Phi}_K\rangle
,\\
\label{eq:spreadqkb}
\langle \Delta I_K^2 \rangle =&  
\langle \Delta \Hat{K}^2 \rangle_S + \langle \Delta \Hat{I}_K^2 \rangle_K 
+ \frac{1}{4} \langle \Delta \Hat{\Phi}_Q^2 \rangle_Q 
+\langle \Delta \Hat{I}_K \Delta \Hat{\Phi}_Q\rangle
,
\end{align}
\end{subequations}
where the indexed brackets indicate averaging over the density matrices 
of system and detectors before the interaction, while the unindexed ones indicate averaging 
over the probability $\Pi(I_Q,I_K)$ given in Eq.~\eqref{eq:probiqik}. 
We notice that Eqs.~\eqref{eq:classqk} are identical in form to Eqs.~\eqref{eq:averqk}. 
This is a consequence of the Ehrenfest theorem, which implies that, for quadratic Hamiltonians, 
the equations of motion for the average values of an observable are identical with the corresponding classical equations.
The formal identity of Eq.~\eqref{eq:probiqik} in the classical and quantum case, however, is a new result, 
going well beyond Ehrenfest theorem. I call this result the strong correspondence principle. 
It amounts to a simple prescription: 
(i) solve the classical equations of motion for the interaction between detectors and system;
(ii) assume an initial joint probability distribution $\Pi(I_Q,\Phi_Q;I_K,\Phi_K)$ for the detectors and $\Pi(Q,K)$ for the system; 
(iii) find the joint probability of observing outcomes $I_Q,I_K$ in terms of the initial probabilities; 
(iv) substitute the classical probability distributions with the Wigner quasi-probability ones.

Let us consider the state of the system after the detection, conditioned on the fact that the readout 
of the detectors was $I_Q,I_K$. 
Rather than working with the density matrix, I consider 
the conditional quasi-characteristic function of the system. 
The unnormalized conditional quasi-charateristic function $\Pi(I_Q,I_K) Z^W_{S}(q,k|I_Q,I_K)$ can be found 
from Eqs.~\eqref{eq:multimeaskern}  by substituting the kernel in \eqref{eq:multimeaskern1} 
with 
\begin{align}
K=&\int\!dQ\ e^{ikQ} \langle Q+\frac{q}{2}|
\Hat{V}_{+} \rho_S
\Hat{V}_{-}^\dagger  
| Q-\frac{q}{2}\rangle .
\end{align}  
This readily gives 
\begin{align*}
& Z^W_{S}(q,k|I_Q,I_K)=\int d\chi_Q d\chi_K \ 
\frac{\exp{\left[-i\left( \chi_Q I_Q+\chi_K I_K\right)\right]}}{\Pi(I_Q,I_K)} 
\\
&\times Z^W_{S}(q+\chi_K,k+\chi_Q) 
Z^W_{det}(\chi_Q,q+\frac{\chi_K}{2};\chi_K,-k-\frac{\chi_Q}{2})
\end{align*}
and the corresponding quasi-probability distribution is 
\begin{widetext}
\begin{align}
\label{eq:classbayes}
& \Pi^W_{S}(K,Q|I_Q,I_K)=\int d\Phi_Q d\Phi_K 
\frac{\Pi^W_{S}(K-\Phi_Q,Q+\Phi_K) 
\Pi^W_{det}(I_Q-Q-\Phi_K/2,\Phi_Q;I_K-K+\Phi_Q/2,\Phi_K)}
{\Pi(I_Q,I_K)},
\end{align}
\end{widetext}
or 
\begin{align}
\nonumber
&\Pi(I_Q,I_K) \Pi^W_{S}(K,Q|I_Q,I_K)=\int dQ' dK' 
\Pi^W_{S}(K',Q') \\
\label{eq:classbayes2}
&\times 
\Pi^W_{det}(I_Q\!-\!\!\frac{Q+Q'}{2},K\!\!-\!K';I_K\!-\!\!\frac{K+K'}{2},Q'\!\!-\!Q).
\end{align}

It has a simple classical interpretation: according to Bayes' theorem, the conditional probability of finding the system 
with values $Q,K$, given that the detectors gave the output $I_Q,I_K$ satisfies 
\begin{equation}
\Pi(I_Q,I_K) \Pi_S(Q,K|I_Q,I_K)= \Pi(Q,K,I_Q,I_K)
\end{equation}
The classical joint probability $\Pi(Q,K,I_Q,I_K)$ can be derived from the classical equations of motion 
Eqs.~\eqref{eq:classqk} 
through the following reasoning: for given $\Phi_Q,\Phi_K$, the values of the system before the interaction must be 
$Q+\Phi_K,K-\Phi_Q$; this happens with probability $\Pi^W_{S}(K-\Phi_Q,Q+\Phi_K)$. The values of $I_Q,I_K$ 
before the interaction must have been $I'_Q=I_Q-Q-\Phi_K/2,I'_K=I_K-K+\Phi_Q/2$, and arbitrary $\Phi_Q,\Phi_K$; 
this happens with probability
$\Pi_{det}(I_Q-Q-\Phi_K/2,\Phi_Q; I_K-K+\Phi_Q/2,\Phi_K)$. 
Integrating over all possible values $\Phi_Q,\Phi_K$ gives Eq.~\eqref{eq:classbayes}. 
Thus we have a further application of the strong correspondence principle: one could derive the joint conditional 
probability through the classical reasoning, and then substitute in the formulas the Wigner quasi-probabilities distributions 
of detectors and system for the positive defined classical probabilities.

We notice that, for general preparation of the detectors, the conditional state of the system depends on its initial state, 
and it is not a gaussian, contrary to what was concluded in \cite{She66} applying the principle of maximum entropy. 

In conclusion, I have demonstrated a rich correspondence between classical and quantum mechanics: 
not only do the average values of an observable obey the classical equations of motion, as established 
by Ehrenfest's theorem, but the full joint probability of the outcomes has the same expression in the classical 
as in the quantum case in terms of the initial probability distributions of the dynamical variables of system and detectors, 
provided that one substitutes the classical joint probabilities with Wigner quasi-probabilities. 
Due to the uncertainty relations, the Wigner quasi-probabilities come in such combinations that they give rise to 
a positive probability distribution. A fecund concept was also introduced, that of quasi-characteristic function, 
in terms of which the characteristic function of the joint outcomes has a remarkably simple expression 
[see Eq. \eqref{eq:charfuncqk}]. From this, one can conclude that detectors 
contribute with an additive term to the cumulants of all orders.
The strong correspondence between the classical and the quantum case was shown to hold also for the 
determination of the conditional state of the system after the measurement.


\begin{thebibliography}{99}

\bibitem{Arthurs65}
E.~Arthurs and J.~L. Kelly, 
Bell Syst. Tech. J.
{\bf 44}, 725 (1965).

\bibitem{Braunstein91}
S.L. Braunstein, C.M. Caves, and G.J. Milburn, Phys. Rev. A \textbf{43},
1153 (1991).
\bibitem{Stenholm92}
S.~Stenholm,
Ann. Phys.\textbf{218},
233 (1992).


\bibitem{Husimi}
K. Husimi, Proc. Phys. Math Soc. Jpn. {\bf 22}, 264 (1940).

\bibitem{Arthurs88}
E. Arthurs and M.S. Goodman. Phys. Rev. Lett. {\bf 60}, 2447 (1988).



\bibitem{uncertainty90s}
M.G. Raymer, Am. J. Phys. {\bf 62}, 986 (1994);
U. Leonhardt, B. B\"{o}hmer, and H. Paul, Opt. Comm. {\bf 119}, 296 (1995); 
D.M. Appleby, J. Phys. A {\bf 31}, 6419 (1998); 
D.~M. Appleby, Int. J. Theor. Phys. \textbf{37}, 1491 (1998); 
D.~M. Appleby, Int. J. Theor. Phys. \textbf{38}, 807 (1999).

\bibitem{uncertainty00s}
G.M. D'Ariano, Fortschr. Phys. {\bf 51}, 318  (2003); 
M. Ozawa, Phys. Rev. A {\bf 67}, 042105 (2003);
M.J.W. Hall, Phys. Rev. A {\bf 69}, 052113 (2004);
E. Andersson, S. M. Barnett, and A, Aspect, Phys. Rev. A {\bf 72}, 042104 (2005);
Yu.I. Vorontsov, Phys. Usp. {\bf 48}, 999 (2005);
P. Busch, T. Heinonen, and P. Lahti, Phys. Rep.{\bf 452}, 155  (2007). 


\bibitem{She66}
C.~Y. She and H.~Heffner,
 Phys.Rev. \textbf{152},
1103 (1966).


\bibitem{vonneumann} 
J. von Neumann, \emph{Mathematische Grundlagen der Quantenmechanik},
Springer, Berlin (1932) [Engl. transl. \emph{Mathematical
Foundations of Quantum Mechanics}, Princeton University
Press (1955)].

\bibitem{Walker} 
N.G. Walker and J.E. Carroll, Electron. Lett. {\bf 20} (1984) 981;
Opt. Quantum Electron. {\bf 18} (1986) 355;
N.G. Walker, J. Mod. Optics {\bf 34} (1987) 15.

\bibitem{Noh} J.W. Noh, A. Foug\`{e}res and L. Mandel, Phys. Rev. Lett. {\bf 67}
(1991) 1426; Phys. Rev. A {\bf 45} (1992) 424; A {\bf 46} (1992) 2840.

\bibitem{Braginsky}
V. B. Braginsky and F. Ya. Khalili, \emph{Quantum measurement}, Cambridge
University Press, Cambridge (1992).

\end{thebibliography}
\end{document}